\documentclass[10pt,letterpaper,twocolumn]{article} %% two column, final layout

\usepackage{ol2}
\usepackage{graphicx}
\usepackage{subfigure}
\usepackage{amsmath,amssymb}
\usepackage{txfonts}

\newcommand{\tr}{\textrm} % Roman math text
\newcommand{\mb}{\mathbf} % Bold math text

\begin{document}

\allowdisplaybreaks

\twocolumn[ %% activate for two-column option

\title{On the validity of the paraxial approximation for electron acceleration\\ with radially polarized laser beams}

\author{Vincent Marceau$^{1,*}$, Charles Varin$^{2}$, and Michel Pich{\'e}$^{1}$}

\address{$^{1}$Centre d'Optique, Photonique et Laser, Universit{\'e} Laval, 2375 de la Terrasse, Qu{\'e}bec (QC), G1V 0A6, Canada\\
$^{2}$Department of Physics, University of Ottawa, 150 Louis Pasteur, Ottawa (ON), K1N 6N5, Canada\\
$^*$Corresponding author: vincent.marceau.2@ulaval.ca
}

\begin{abstract}
In the study of laser-driven electron acceleration, it has become customary to work within the framework of paraxial wave optics. Using an exact solution to the Helmholtz equation as well as its paraxial counterpart, we perform numerical simulations of electron acceleration with a high-power TM$_{01}$ beam. For beam waist sizes at which the paraxial approximation was previously recognized valid, we highlight significant differences in the angular divergence and energy distribution of the electron bunches produced by the exact and the paraxial solutions. Our results demonstrate that extra care has to be taken when working under the paraxial approximation in the context of electron acceleration with radially polarized laser beams. 
\end{abstract}


 ] %% activate for two-column option

Electron acceleration driven by high-power lasers has attracted much effort, since it could lead to the development of compact electron accelerators. Such devices are expected to be useful in many applications, ranging from coherent X-ray generation~\cite{nakajima08_naturephys,schlenvoigt08_naturephys} to electron diffraction imaging~\cite{baum07_pnas}. In laser-driven particle acceleration, the challenge is to find an efficient way to apply the electromagnetic field. Among the many proposed methods, the use of radially polarized laser beams (RPLBs) for electron acceleration in vacuum has been the object of many investigations~\cite{salamin06_pra,varin05_pre,varin06_pre,salamin07_optlett,fortin10_jpb,wong10_optexpress,singh11_prstab,marceau12_optlett}. In this scheme, which was recently demonstrated experimentally~\cite{payeur12_apl}, electrons experience sub-cycle acceleration from the longitudinal electric field component at the center of an ultra-intense TM$_{01}$ beam. This mechanism induces a strong longitudinal compression effect that could theoretically lead to the production of attosecond electron bunches~\cite{varin06_pre,karmakar07_lpb}. 

Many studies that have investigated on-axis~\cite{varin05_pre,salamin06_pra,fortin10_jpb,wong10_optexpress} and off-axis~\cite{varin06_pre,singh11_prstab} electron acceleration in RPLBs are based on the usual \textit{paraxial approximation}. In these analyses, it is generally assumed that the paraxial approximation is valid whenever $k_0 z_R \gtrsim 125$ (or $w_0/\lambda_0 \gtrsim 2.5$), where $z_R$ and $w_0$ are respectively the Rayleigh range and beam waist size, and $\lambda_0 = 2\pi/k_0$ is the dominant wavelength of the laser. 

In this Letter, we question the validity of the paraxial approximation in the above-mentioned parameter regime. Using a rigorous solution to Helmholtz equation for a TM$_{01}$ beam as well as its paraxial counterpart, we highlight significant differences in the electron acceleration dynamics predicted by the exact and the paraxial fields. The exact fields are used to investigate the origin and importance of these differences.

A nonparaxial TM$_{01}$ beam propagating along the positive $z$ axis with its beam waist at $z=0$ is described in complex notation (i.e., $E=\tr{Re}\{\tilde{E}e^{j(\omega_0 t-\phi_0)}\}$, where $\omega_0=ck_0$ and $\phi_0$ are the beam frequency and phase constant) by the following field components in cylindrical coordinates $(r,\phi,z)$~\cite{sheppard99_ol,april08b_optlett}:
\begin{align}
&\tilde{E}_r = -j E_0 e^{-k_0 a} j_2 (k_0 \tilde{R}) \sin \tilde{\theta} \cos \tilde{\theta} \label{eq:exactEr} \\
&\tilde{E}_z  = -\tfrac{2}{3}jE_0 e^{-k_0 a}\left[j_0 (k_0 \tilde{R}) + j_2(k_0 \tilde{R}) P_2(\cos \tilde{\theta}) \right] \label{eq:exactEz}  \\
&\tilde{B}_\phi = B_0 e^{-k_0 a} j_1 (k_0 \tilde{R}) \sin \tilde{\theta} \label{eq:exactHphi} \, .
\end{align}
Here $E_0=c B_0$ are amplitude parameters, $k_0$ is the beam wavenumber, $\tilde{R}=[r^2+(z+ja)^2]^{1/2}$ is the complex radius, $\cos \tilde{\theta}=(z+ja)/\tilde{R}$ defines the complex angle $\tilde{\theta}$,  $j_n(\cdot)$ is the $n$-th order spherical Bessel function of the first kind, and $P_2(\cdot)$ is the Legendre polynomial of degree 2. The parameter $a$ is a real and positive constant called the confocal parameter. The latter may be used to characterize the degree of paraxiality of the beam since it is monotonically related to the Rayleigh range and beam waist size by the relation $z_R = k_0 w_0^2/2 = [\sqrt{1+(k_0a)^2}-1]/k_0$. The power carried along the $z$ axis by a nonparaxial TM$_{01}$ beam is~\cite{april10_josaa}
\begin{align}
\!\!\! P \!=\! \frac{P_0 e^{-2k_0 a}}{ k_0^3 a^3} \left[2k_0 a \sinh(2k_0 a)\!-\!\cosh(2k_0 a) \!+\! 1 \!-\! 2k_0^2 a^2 \right],
\end{align}
where $P_0 = \pi |E_0|^2/8 \eta_0 k_0^2$. Note that the fields described in Eqs.~\eqref{eq:exactEr}--\eqref{eq:exactHphi} represent a rigorous solution to Helmholtz equation. We will thus refer to them as the \textit{exact} TM$_{01}$ \textit{fields}.

In the paraxial limit, namely when $k_0 a \gg 1$, the fields \eqref{eq:exactEr}--\eqref{eq:exactHphi} can be expanded as power series of the parameter $\delta = 1/k_0 a$. Using the normalized coordinates $\rho =  r(2a/k_0)^{-1/2}$ and $\zeta = z/a$ (since $z_R=k_0w_0^2/2 \approx a$), we find, up to terms of order $\delta^3$:
\begin{align}
&\begin{aligned}
\tilde{E}_r = - \tfrac{1}{\sqrt{2}}E_0  \Big[ \rho f^2\delta^{3/2} - \big( 3j\rho f^3\!+\!3&\rho^3 f^4 \!-\!\tfrac{j}{2}\rho^5 f^5 \big)\delta^{5/2}\\ 
 & \ \ + \mathcal{O} (\delta^{7/2}) \Big] \exp(j\Psi)
\end{aligned}\label{eq:corrEr}\\
&\begin{aligned}
\tilde{E}_z &= jE_0 \Big[\big( f^2 \!-\! j \rho^2 f^3 \big)\delta^2 \\
&- \big(jf^3 \!+\! 5\rho^2 f^4 \!-\! \tfrac{7j}{2} \rho^4 f^5 \!-\! \tfrac{1}{2}\rho^6 f^6 \big)\delta^3+ \mathcal{O} (\delta^{4}) \Big] \exp(j\Psi)
\end{aligned}\label{eq:corrEz}	\\
&\begin{aligned}
\tilde{B}_\phi = -\tfrac{1}{\sqrt{2}}B_0 \Big[ \rho f^2\delta^{3/2} - \big( j\rho f^3\!+\!2&\rho^3 f^4 \!-\!\tfrac{j}{2}\rho^5 f^5 \big)\delta^{5/2}\\ 
 & \ \ + \mathcal{O} (\delta^{7/2}) \Big] \exp(j\Psi)  \ ,
\end{aligned}\label{eq:corrHphi}
\end{align}
where $f=1/(\zeta+j)$ and $\Psi = - (\zeta/\delta + \rho^2f )$. The lowest-order terms of Eqs.~\eqref{eq:corrEr}--\eqref{eq:corrHphi} correspond to the well-known \textit{paraxial} TM$_{01}$ \textit{fields}, which are commonly used to analyze electron acceleration in RPLBs~\cite{varin05_pre,salamin06_pra,varin06_pre,fortin10_jpb,wong10_optexpress,singh11_prstab},
\begin{align}
&\tilde{E}_r^{(0)} = -\tfrac{1}{\sqrt{2}}\, E_0 \rho f^2\delta^{3/2} \exp(j\Psi) = c B_\phi^{(0)}\label{eq:paraEr}\\
&\tilde{E}_z^{(0)} = jE_0 \big( f^2 - j \rho^2 f^3 \big)\,\delta^2 \exp(j\Psi) \label{eq:paraEz}\ .
\end{align}

To simulate electron acceleration in RPLBs, we numerically integrate the Newton-Lorentz equations:
\begin{align}
\frac{d\mb{r}}{dt} = \mb{v} \ ,\ \ \ \frac{d\mb{v}}{dt} = -\frac{e}{\gamma m_e}\left[\mb{E}+ \mb{v}\times\mb{B} - \frac{\mb{v}}{c^2}\left(\mb{v}\cdot\mb{E}\right)\right] ,
\end{align}
where $e$, $m_e$, $\mb{r}$, $\mb{v}$ are the electron's charge, mass, position, velocity, respectively, and $\gamma=(1-|\mb{v}|^2/c^2)^{-1/2}$. We also suppose that the laser beam is pulsed, which we model by multiplying $\mb{E}$ and $\mb{B}$ by $\tr{sech}(\xi/\xi_0)$, where $\xi = \omega_0 t - k_0 z$. This is to ensure that the fields satisfy Maxwell's equation in the limit $\xi_0 \gg 1$ for any value of the phase $\xi$~\cite{wong10_optexpress}. Furthermore, we use $\lambda_0 = 800$ nm, although it can be readily shown that the results are scalable to any value of $\lambda_0$~\cite{wong10_optexpress,marceau12_optlett}.

\begin{figure}[!t]
\centering
\includegraphics[width=\columnwidth]{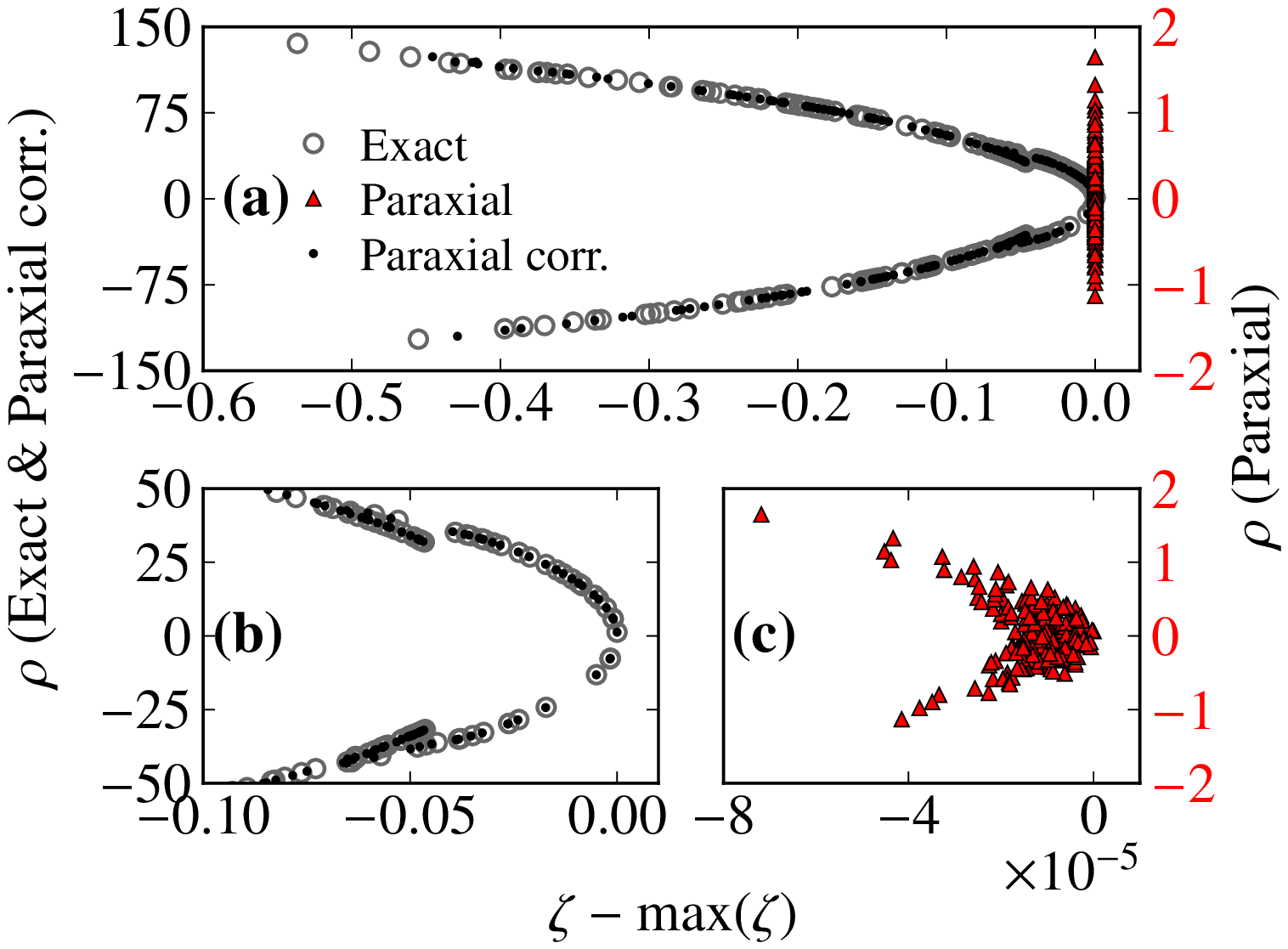} \\ 
\includegraphics[width=\columnwidth]{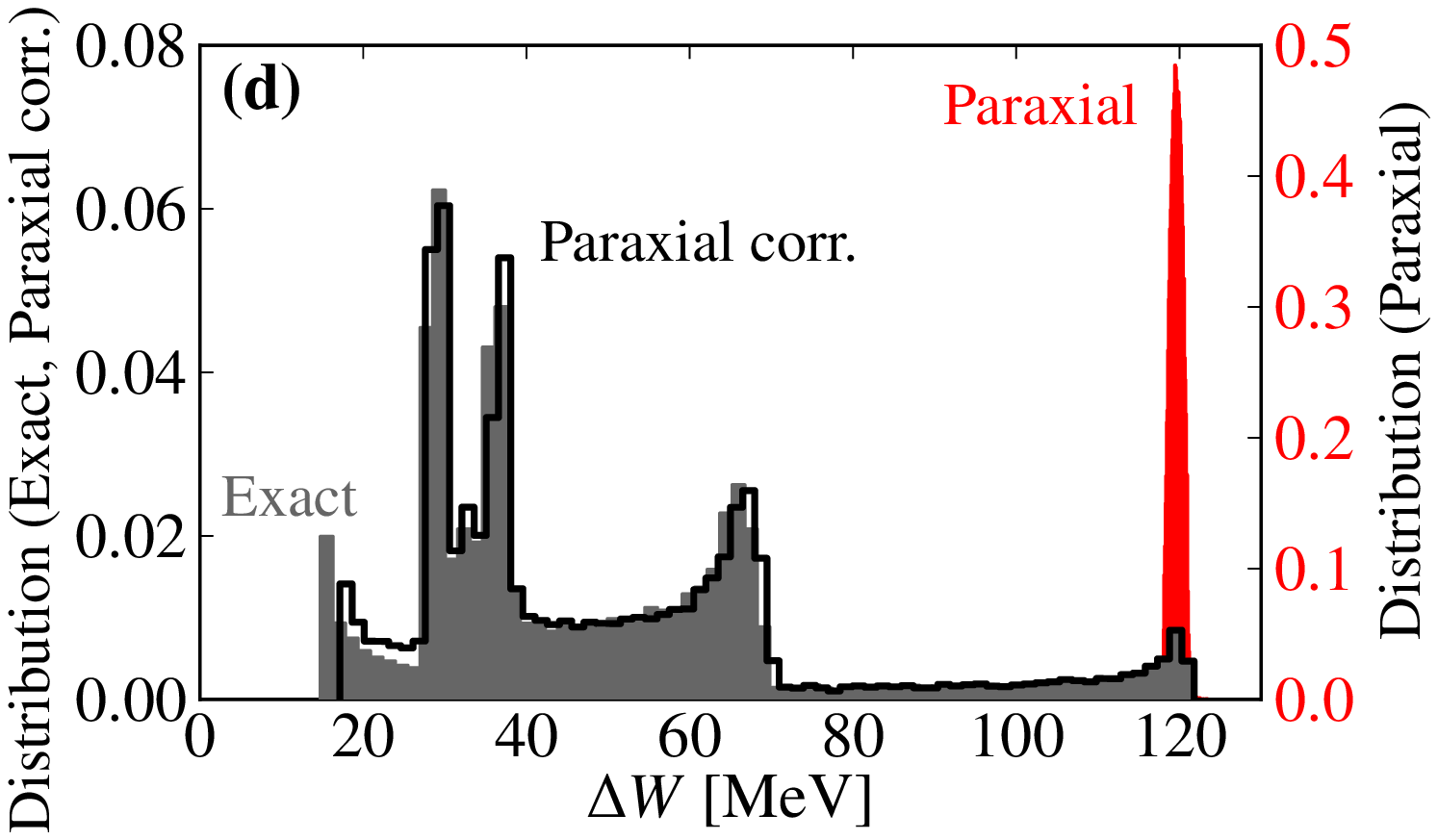} \\ 
\includegraphics[width=\columnwidth]{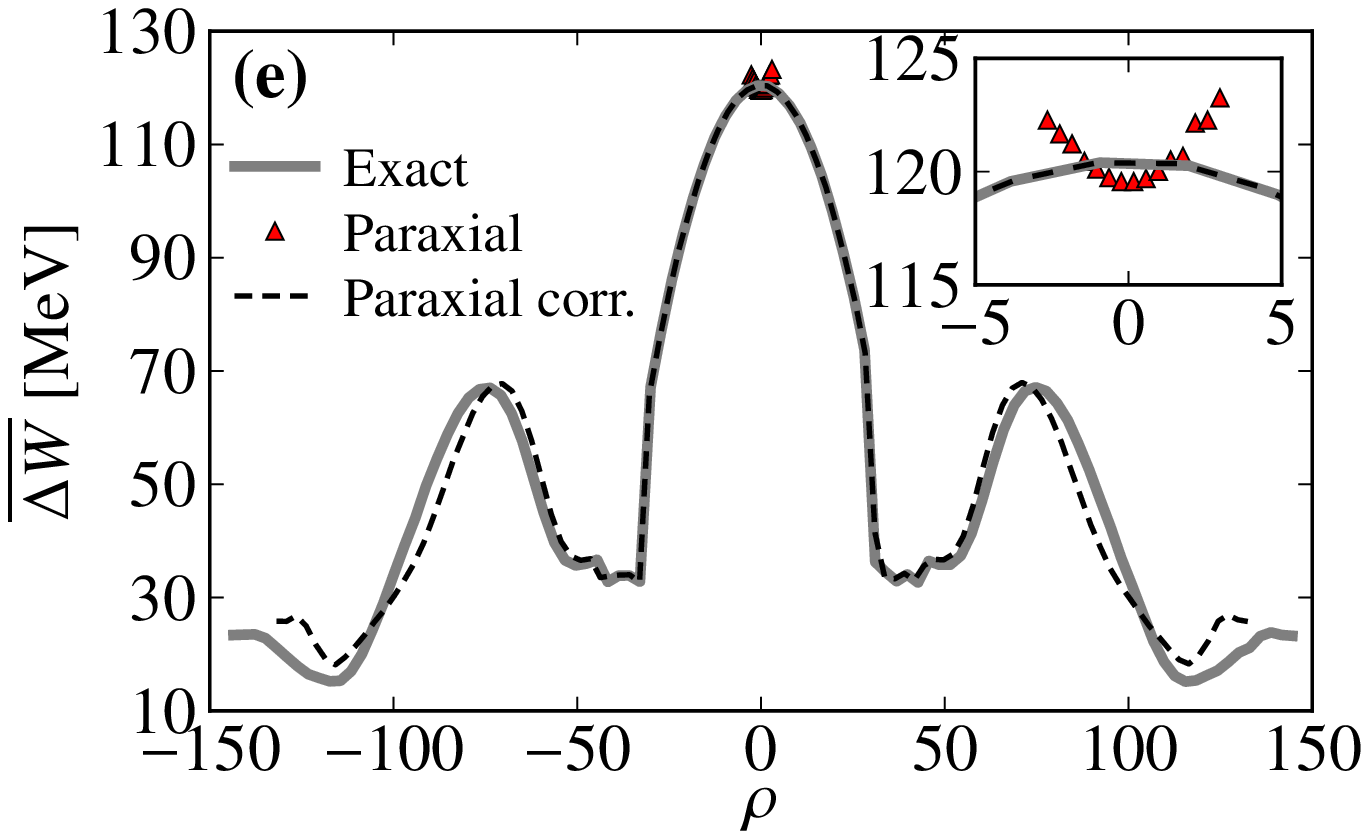} \\ 
\caption{(Color online) Electron bunch accelerated by a TM$_{01}$ pulsed beam with $P=10^{15}$ W, $k_0 a = 500$, $\xi_0=14.21$, $\phi_0=\pi$. The simulation was performed with: (i) the exact fields [Eqs.~\eqref{eq:exactEr}--\eqref{eq:exactHphi}], (ii) the paraxial fields [Eqs.~\eqref{eq:paraEr}--\eqref{eq:paraEz}], and (iii) the corrected paraxial fields [Eqs.~\eqref{eq:corrEr}--\eqref{eq:corrHphi} up to $\mathcal{O}(\delta^3)$]. (a) Bunch snapshot and (b)--(c) close-up view of the front end; (d) energy gain distribution; (e) average energy gain versus final radial coordinate. The electrons are initially at rest outside the laser pulse in the $(r,z)$ plane and distributed randomly according to a Gaussian distribution centered at the origin with $\sigma_r = \sigma_z = \lambda_0/10$. The results are computed 15 ps after the passage of the beam at $z=0$. Only $N=200$ electrons are shown in (a)--(c), while $N=50000$ different initial conditions were used to obtain the results in (d) and (e). \label{fig:bunches}}
\end{figure}

We consider a cloud of electrons initially at rest in the $(r,z)$ plane outside the laser pulse. The initial position of the electrons are drawn randomly from a two-dimensional Gaussian distribution centered at the origin with standard deviation $\sigma_r = \sigma_z = \lambda_0/10$. The electrons are accelerated by a pulsed TM$_{01}$ beam with $k_0 a = 500$ (which corresponds to $w_0 \approx 5\lambda_0$), a value generally considered well inside the paraxial regime. Space-charge effects are neglected; each trajectory is computed independently. The relativistic electron bunch produced after performing the numerical simulation with the exact and the paraxial TM$_{01}$ fields from the same initial conditions are shown in Fig.~\ref{fig:bunches}. A comparison of the datasets in Fig.~\ref{fig:bunches}(a)--(c) immediately shows an enormous difference in the bunch transverse extent; the angular divergence of the bunch accelerated by the exact fields is about 100 times larger than the bunch accelerated by the paraxial fields. Moreover, the electron bunch energy gain distribution is very different from one case to the other, as shown in Fig.~\ref{fig:bunches}(d). The average energy gain obtained with the paraxial fields is however close to the energy gain near the optical axis with the exact fields, as shown in Fig.~\ref{fig:bunches}(e). This agrees with previous results reported in~\cite{marceau12_optlett} for on-axis acceleration. Since the near-axis electrodynamics is similar in both cases, the strong longitudinal compression predicted by the paraxial fields is also observed with the exact fields.

To understand why such important discrepancies between the exact and the paraxial cases arise for off-axis electrons, it is instructive to look at the equations of motion. When electrons close to the optical axis interact with the laser beam, they are primarily accelerated in the positive $z$ direction by the longitudinal electric field. For an electron with $v_r \ll v_z$, the equation governing the radial velocity is approximately
\begin{align}
\frac{dv_r}{dt} \approx \frac{-e}{\gamma m_e} \left(E_r - v_z B_\phi \right) \ . \label{eq:Lorentz}
\end{align}
From Eq.~\eqref{eq:paraEr}, we see that the paraxial $\tilde{E}_r$ and $\tilde{B}_\phi$ field components are perfectly symmetric, i.e., $\tilde{E}_r^{(0)} = c\tilde{B}_\phi^{(0)}$. Therefore, according to Eq.~\eqref{eq:Lorentz}, an electron travelling primarily along the optical axis at a relativistic velocity will feel a quasi-null force in the radial direction. This explains the observation of electron bunches with very narrow transverse extent when performing the simulations with the paraxial fields. 

\begin{figure}[!t]
\centering
\includegraphics[width=1\columnwidth]{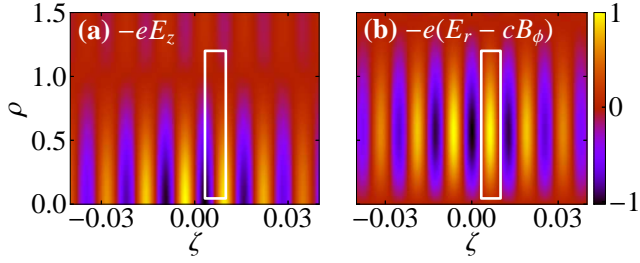} \\ 
\caption{(Color online) Normalized amplitude of (a) $-eE_z$ and (b) $-e(E_r - cB_\phi)$ at $t=0$ (exact fields). During sub-cycle acceleration, electrons will spend most of their time between a maximum of $-eE_z$ and the minimum located behind [an example is shown by the rectangle in (a) and (b)]. In this phase configuration, electrons travel at $v_z \approx c$ and feel, according to (b), a radial force directed outward. Beam parameters are the same as in Fig.~\ref{fig:bunches}.  \label{fig:fields}}
\end{figure}

However, this argument does not hold in the exact case. According to Eqs.~\eqref{eq:exactEr} and \eqref{eq:exactHphi}, the exact $\tilde{E}_r$ and $\tilde{B}_\phi$ field components are not perfectly symmetric. The perturbative series \eqref{eq:corrEr} and \eqref{eq:corrHphi} show that the symmetry between $\tilde{E}_r$ and $\tilde{B}_\phi$ is broken as soon as the first nonparaxial correction is introduced. Since $E_r$ and $B_\phi$ reach their maximum value at $\rho \approx [(1+\zeta^2)/2]^{1/2}$, the ratio of the maximum value of the radial field term in Eq.~\eqref{eq:Lorentz}, $|E_r - cB_\phi|$, and the accelerating field, $|E_z|$, scales as $ \delta^{1/2}(1+\zeta^2)^{1/2}$. The radial force component may therefore not be neglected when $k_0 a \gtrsim 100$, and its relative importance becomes greater as $\zeta$ increases. Moreover, Fig.~\ref{fig:fields} demonstrates that the phase configuration in which electrons spend most of their time during sub-cycle acceleration is associated with a radial force directed outward. This explains why the angular divergence of the bunch accelerated by the exact solution is larger instead of smaller.  

To verify explicitly if the discrepancy between the electron bunches in Fig.~\ref{fig:bunches} originates from the articifial symmetry between the paraxial $\tilde{E}^{(0)}_r$ and $\tilde{B}^{(0)}_\phi$ field components, we have performed the same numerical simulation with the \textit{corrected paraxial fields}, i.e., the paraxial fields with the first nonparaxial correction to each component [Eqs.~\eqref{eq:corrEr}--\eqref{eq:corrHphi} up to $\mathcal{O}(\delta^3)$]. The results obtained are shown in Fig.~\ref{fig:bunches}. We immediately see a much better agreement with the exact case. While the final positions of the electrons do not match perfectly, the spatial configuration of the bunches and their energy gain distribution are much more similar.  Note that the small differences in the energy gain distribution can be understood from Fig.~\ref{fig:bunches}(e). Indeed, we see that the corrected paraxial results get less accurate as $\rho$ increases. This is a consequence of the fact that the magnitude of the nonparaxial corrections increases as we move further from the optical axis. Taking into account the first nonparaxial correction to $\tilde{E}_r$ and $\tilde{B}_\phi$ thus only offers a limited solution for electrons far from the optical axis.

Despite the fact that the validity of the paraxial approximation was never fully adressed in the literature for $k_0z_R$ values well above 100, an order of magnitude comparison may be performed with existing results. In~\cite{karmakar07_lpb}, an RPLB with $w_0 = 3\lambda_0$ and second-order field accuracy was used to accelerate electrons from a target of size comparable to the initial conditions used here. The resulting electron bunch angular divergence was reported to be approximately $\Delta \theta \approx 3^\circ$. With the parameters used in this Letter, we obtain $\Delta \theta \approx 5^\circ$ with the exact fields, which is of the same order of magnitude as in~\cite{karmakar07_lpb}, compared to $\Delta \theta <  0.1^\circ$ for the paraxial fields.

In conclusion, in a parameter regime where the paraxial approximation was previously considered valid, we have highlighted significant differences between the properties of electron bunches accelerated by paraxial and exact TM$_{01}$ beams. These differences originate from the symmetry between the paraxial $\tilde{E}_r^{(0)}$ and $\tilde{B}_\phi^{(0)}$ fields. This artificial symmetry is broken as soon as the first nonparaxial corrections to the electromagnetic field are taken into account, which allows to obtain more accurate results. The considerations presented in this Letter are also believed to apply to ultrashort pulses, since the relation $\tilde{E}_r = c\tilde{B}_\phi$ always holds for the paraxial TM$_{01}$ fields, regardless of the pulse duration~\cite{varin05_pre}. Our study thus advocates that special care has to be taken when working under the paraxial approximation in the context of electron acceleration in RPLBs. It should be reminded that under relativistic conditions, nonparaxial field corrections may always yield major differences in the trajectories of off-axis electrons, even for very large values of $k_0 a$. Solutions as accurate as possible, ideally exact, should thus be used. 

This research was supported by the Natural Sciences and Engineering Research Council of Canada (NSERC), the Canadian Institute for Photonic Innovations (CIPI), and Calcul Qu\'ebec Universit\'e Laval (computational resources).

%\bibliographystyle{ol}
%\bibliography{./marceau_biblio.bib}

\begin{thebibliography}{10}
\newcommand{\enquote}[1]{``#1''}

\bibitem{nakajima08_naturephys}
K.~Nakajima, Nature Phys. \textbf{4}, 92 (2008).

\bibitem{schlenvoigt08_naturephys}
H.-P.~Schlenvoigt, K.~Haupt, A.~Debus, F.~Budde, O.~J\"ackel, S.~Pfotenhauer, H.~Schwoerer, E.~Rohwer, J.~G.~Gallacher, E.~Brunetti, R.~P.~Shanks, S.~M.~Wiggins, and D.~A.~Jaroszynski, Nature Phys. \textbf{4}, 130 (2008).

\bibitem{baum07_pnas}
P.~Baum and A.~H. Zewail, Proc. Natl. Acad. Sci. USA \textbf{104}, 18409
  (2007).

\bibitem{salamin06_pra}
Y.~I. Salamin, Phys. Rev. A \textbf{73}, 043402 (2006).

\bibitem{varin05_pre}
C.~Varin, M.~Pich{\'e}, and M.~A. Porras, Phys. Rev. E \textbf{71}, 026603
  (2005).

\bibitem{varin06_pre}
C.~Varin and M.~Pich{\'e}, Phys. Rev. E \textbf{74}, 045602(R) (2006).

\bibitem{salamin07_optlett}
Y.~I. Salamin, Opt. Lett. \textbf{32}, 90 (2007).

\bibitem{fortin10_jpb}
P.-L. Fortin, M.~Pich{\'e}, and C.~Varin, J. Phys. B \textbf{43}, 025401
  (2010).

\bibitem{wong10_optexpress}
L.~J. Wong and F.~X. K{\"a}rtner, Opt. Express \textbf{18}, 25035 (2010).

\bibitem{singh11_prstab}
K.~P. Singh and M.~Kumar, Phys. Rev. ST -- Accel. Beams \textbf{14}, 030401
  (2011).

\bibitem{marceau12_optlett}
V.~Marceau, A.~April, and M.~Pich{\'e}, Opt. Lett. \textbf{37}, 2442 (2012).

\bibitem{payeur12_apl}
S.~Payeur, S.~Fourmaux, B.~E. Schmidt, J.-P. MacLean, C.~Tchervenkov,
  F.~L{\'e}gar{\'e}, M.~Pich{\'e}, and J.-C. Kieffer, Appl. Phys. Lett.
  \textbf{101}, 041105 (2012).

\bibitem{karmakar07_lpb}
A.~Karmakar and A.~Pukhov, Laser Part. Beams \textbf{25}, 371 (2007).

\bibitem{sheppard99_ol}
C.~J.~R. Sheppard and S.~Saghafi, Opt. Lett. \textbf{24}, 1543 (1999).

\bibitem{april08b_optlett}
A.~April, Opt. Lett. \textbf{33}, 1563 (2008).

\bibitem{april10_josaa}
A.~April, J. Opt. Soc. Am. A \textbf{27}, 76 (2010).

\end{thebibliography}

\end{document}